\documentclass[conference]{IEEEtran}

\usepackage{tabularx}
\begin{document}

\title{A comprehensible analysis of the efficacy of Ensemble Models for Bug Prediction}

\author{\IEEEauthorblockN{Ingrid Marçal}
\IEEEauthorblockA{Department of Computer Science\\
São Paulo State University\\
Presidente Prudente, SP\\
Email: ingrid.marcal@unesp.br}
\and
\IEEEauthorblockN{Rogério Eduardo Garcia}
\IEEEauthorblockA{Department of Computer Science\\
São Paulo State University\\
Presidente Prudente, SP\\
Email: rogerio.garcia@unesp.br}
}

\maketitle

\begin{abstract}
The correctness of software systems is vital for their effective operation. It makes discovering and fixing software bugs an important development task. The increasing use of Artificial Intelligence (AI) techniques in Software Engineering led to the development of a number of techniques that can assist software developers in identifying potential bugs in code. In this paper, we present a comprehensible comparison and analysis of the efficacy of two AI-based approaches, namely single AI models and ensemble AI models, for predicting the probability of a Java class being buggy. We used two open-source Apache Commons Project's Java components for training and evaluating the models. Our experimental findings indicate that the ensemble of AI models can outperform the results of applying individual AI models. We also offer insight into the factors that contribute to the enhanced performance of the ensemble AI model. The presented results demonstrate the potential of using ensemble AI models to enhance bug prediction results, which could ultimately result in more reliable software systems. 
\end{abstract}

\maketitle

\section{Introduction}


Bugs have been an essential subject of research in the field of software engineering for a long time, mainly due to their importance as an indicator of software quality \cite{Ma_2020}. A significant research contribution in this subject is the development of bug prediction approaches that aim to determine the areas of the source code that are more likely to be buggy \cite{Chun_2014, Habibi_2018}. The results obtained from applying bug prediction can be subsequently used to guide refactoring, code review, and bug-fixing activities.

Despite its importance, identifying potential bugs in code can be a challenging task \cite{Ishani_2015, Kamei_2016, Bhandari_2022}. With the increasing use of AI techniques in Software Engineering, researchers have explored the use of machine learning models to predict bugs in code \cite{Chun_2014, Habibi_2018, Soe_2018, Rahim_2021, Aleithan_2021, Shen_2022}.

Several AI-based approaches have been proposed for predicting software bugs, including single AI models \cite{Chun_2014, Habibi_2018, Soe_2018, Rahim_2021, Aleithan_2021, Shen_2022} and ensemble AI models \cite{Santosh_2017, Alazba_2022, Johnson_2022}. While single AI models have been found to be effective in predicting software bugs, ensemble AI models have the potential to improve bug prediction results further.

Single models refer to machine learning models that are trained using a single algorithm. These models can be used to make predictions or classifications based on input data, and they are commonly used in various applications, including image recognition, natural language processing, and fraud detection. Examples of single AI models include decision trees, neural networks, and support vector machines.

Ensemble models refer to machine learning models that are composed of multiple base models, and each trained using a different algorithm. The results of these base models are combined to make a final prediction, which can be more accurate and robust than the prediction made by a single model. Ensemble models are also widely used in various applications, for example, recommendation systems.

According to Nisbet et al. \cite{Nisbet_2017}, one of the motivations for using ensemble models is to reduce the generalization error of the predictions. When the selected base models are diverse and independent, it is possible to minimize prediction errors with the combination of them.

In this paper, we aim to compare the effectiveness of single AI models and ensemble AI models for predicting the probability of a Java class being buggy. Specifically, we use two open-source Apache Commons Project's Java components for training and evaluating bug prediction models. We also investigate the factors that contribute to the enhanced performance of the ensemble AI model.

In our investigation, we train and evaluate both single AI models and ensemble AI models. We use precision, recall, AUC, and F1-score as evaluation metrics to measure the performance of the models in identifying buggy-prone Java files in each project.

The presented results demonstrate the potential of using ensemble AI models to enhance bug prediction results, which could ultimately result in more reliable software systems. This paper contributes to the existing literature by comprehensively comparing single AI models and ensemble AI models for bug prediction in Java components. By identifying the factors contributing to the enhanced performance of ensemble AI models, our findings provide insights into how to improve the efficacy of bug prediction models in the future.

\section{Background and Related Work}
Bug prediction has emerged as a prominent area of interest within software engineering, owing to its potential to facilitate the identification of software vulnerabilities. The utilization of machine learning-based models has become increasingly prevalent for bug prediction. This is due to the model's ability to provide precise and accurate evaluations of software characteristics, enabling the prediction of potential bugs.

Initial research in software defect prediction was primarily focused on evaluating various machine learning models to enhance the precision and effectiveness of bug prediction \cite{Jalaj_2022}. Extensive research has been conducted with Bayesian algorithms, comprising a range of statistical models based on Bayes' theorem \cite{Turhan_2009, Dejaeger_2013, Okutan_2014}. These algorithms are Naive Bayes, Bayesian Belief Network (BBN), and Bayesian Network. The Support Vector Machine (SVM) is a popular machine-learning technique that employs similarity measures to predict bugs. The K-means and Hierarchical Clustering algorithms effectively group data based on shared traits. By organizing the data into clusters with high commonality, these algorithms enable the identification of areas of code where software bugs are likely to occur.

Prior research has demonstrated that no single classifier consistently outperforms others, as their effectiveness depends on the specific dataset under consideration \cite{Delgado_2014, Caruana_2006}.

When applied to Bug Prediction, ensemble techniques combine different classifiers to improve bug prediction performance. Tosun et al. \cite{Tosun_2008} conducted an experimental study on the Validation and Voting technique, which involves aggregating the outputs of various classifiers using a specific aggregating function. In this case, the aggregation function relies on a majority voting scheme, whereby a class is classified as buggy if most models predict it. This approach is effective in practice, improving the performance of prediction models \cite{Jacob_2021}. In contrast, the classification task is hindered when most models fail to predict bugginess.

In this work, we aim to investigate the performance of single models instead of ensemble models when applied to bug prediction in open-source systems. We experiment with seven well-known machine learning algorithms, CART \cite{Breiman_1984}, KNN \cite{Cover_1967}, LDA \cite{Fisher_1936}, LR \cite{Hosmer_2005}, NB \cite{Cestnik_1990}, RF \cite{Breiman_2001}, and SVM \cite{Cortes_1995}. In addition, we analyze a voting classifier as an ensemble model prediction strategy \cite{Dietterich_2000}. The purpose of these experiments is to be able to answer the following research questions:

\begin{itemize}
\item \textbf{RQ1}: How does the performance of single models compare regarding defect prediction accuracy for the Apache Commons systems?
\item \textbf{RQ2}: Does the ensemble model outperform the individual models regarding defect prediction accuracy, precision, recall, F1 score, and AUC?
\item \textbf{RQ3}: What are the trade-offs between single and ensemble models for defect prediction?
\end{itemize}

\section{Methodology}

This section describes the methodology we used to execute and compare the outcomes of single-model prediction with those of the ensemble model in the context of the chosen subject systems. We defined a comprehensive methodology encompassing several stages: subject systems selection, data collection, model selection, training, ensemble model construction, and evaluation metrics. These stages are crucial in ensuring the research findings' reliability, validity, and reproducibility.

The choice of subject systems stage involves carefully identifying and selecting relevant systems that align with the research objectives. The data collection stage involves the systematic gathering of metrics from the development history of each subject system, which is then used to train the models. The model selection stage involves identifying and selecting appropriate models best suited to the research objectives. The training stage involves the iterative process of refining the models using training data. The ensemble model construction stage involves the integration of multiple models to improve the accuracy and robustness of the predictions. Finally, the evaluation metrics stage involves using appropriate metrics to assess the models' performance and the research findings' validity.

Overall, we defined this methodology to provide a rigorous and systematic approach to conducting experiments with different bug prediction algorithms. The following sections describe each phase of the defined methodology.

\subsection{Subject Systems}

The availability of an appropriate dataset is crucial for conducting an adequate analysis of defect prediction models. In this study, we chose projects from Apache Commons to generate the necessary datasets. The selected projects are \textsl{commons-cli} and \textsl{commons-compress}.

These Apache Commons projects are well-established software components widely adopted in the open-source community and offer a substantial codebase with extensive historical information (Table  \ref{tab:subjectSystems}). Also, they represent different types of functionality and domains, making them suitable for evaluating the performance of defect prediction models across different software contexts.

The selection systems' diversity allows for exploration by the prediction model of potential variations in defect characteristics, coding styles, and software development practices, which can further enhance its generalizability.

\begin{table}[ht]
\caption{Subject systems approximate size in lines of code and number of revisions.}
\label{tab:subjectSystems}
\begin{tabular}{|l|l|l|}
\hline
\textbf{Subject} & \textbf{Number of Commits} & \textbf{Approximate Size (LOC)} \\ \hline
commons-cli & 1.285 & 11.495 \\ \hline
commons-compress & 4.077 & 232.447 \\ \hline
\end{tabular}
\end{table}

\subsection{Data Collection}
Obtaining datasets that are tailored to the specific needs of one's research can be a challenging task. To address this issue, we have developed our datasets through a meticulous data collection process encompassing a range of software metrics. By developing our datasets, we ensure that their content is comprehensive, accurate, and well-suited to our research needs. In this study, we collected metrics from the software repository of each subject system. The metrics were carefully selected to ensure that they provide relevant information about the systems under investigation and allow answering the defined research questions. The following metrics were collected from the software repository of each subject system in the Apache Commons.

\begin{itemize}
\item \textbf{Number of authors}: The number of authors contributing to a file provides a measure of collaboration and the diversity of expertise involved in its development. Research has shown that defects can be influenced by the characteristics and experience levels of developers \cite{Yu_2021}. Considering this metric, we aim to explore the relationship between team dynamics and defect proneness.
\item \textbf{Age}: The age of a file, measured as the time elapsed since the last commit, reflects its stability and maturity. Older files may have undergone more testing and refinement, potentially reducing their defect-proneness. Conversely, recent modifications may make newer files more prone to defects. Including age as a metric allows us to investigate the impact of file maturity on defect prediction.
\item \textbf{Number of unique changes}: The count of unique changes made to a file captures the level of modification and activity surrounding it. Files that have undergone frequent changes may exhibit higher defect densities, as each change introduces the potential for new defects. By considering this metric, we aim to capture the volatility of files and their relationship with defect occurrence.
\item \textbf{Size}: File size, measured in lines of code, is a widely studied metric in defect prediction. Larger files tend to be more complex and may contain more opportunities for defects to arise. We aim to investigate the relationship between code complexity and defect proneness by including size as a metric.
\item \textbf{Lines added and Lines deleted}: The number of lines added and deleted from a file reflects the modification and maintenance activities surrounding it. Files with a high number of lines added or deleted may undergo significant changes that can introduce defects. Including these metrics allows us to capture the impact of code modifications on defect prediction.
\end{itemize}

By selecting these metrics, we aim to comprehensively understand the factors contributing to defect occurrence in the Apache Commons systems. Their inclusion allows us to explore various dimensions of software development, such as team dynamics, code evolution, complexity, and maintenance activities, and their influence on defect proneness.

The chosen metrics were obtained by executing git commands on each repository and subsequently processing the results to extract the corresponding numeric values. For example, the "number of authors" metric refers to the sum of distinct commit authors of a given file.

The target variable in our dataset corresponds to whether a specific file in a particular commit is classified as buggy or not. To determine the target variable, we utilized the SZZ algorithm \cite{Borg_2019}, which helped identify the files marked as buggy in each commit. During the training phase, the target variable was not used as a feature for model training. However, it was used to validate the results obtained from applying each model, allowing us to assess the accuracy and effectiveness of the models in predicting the buggy files in different commits.

\subsection{Model Selection}
The selection of suitable machine learning models is an important activity in defect prediction research, as it directly influences the precision and efficacy of the prediction results. In this work, we selected various models, which were subsequently assessed for their efficacy in predicting defects within the Apache Commons systems. The models selected for the study are as follows:

\begin{itemize}
\item \textbf{CART (Classification and Regression Trees)}: CART is a decision tree-based model that recursively partitions the data based on feature values to form a tree-like model. It is known for its simplicity, interpretability, and ability to handle categorical and numerical data. By including CART, we can assess the predictive power and interpretability of decision tree-based models in defect prediction \cite{Breiman_1984}.
\item \textbf{KNN (k-Nearest Neighbors)}: KNN is a non-parametric model that classifies instances based on their proximity to labeled instances in the training data. It is particularly suitable for identifying patterns in data without assuming any underlying distribution. By incorporating KNN, we aim to evaluate the performance of instance-based learning approaches in defect prediction \cite{Cover_1967}.
\item \textbf{LDA (Linear Discriminant Analysis)}: LDA is a statistical model that aims to find linear combinations of features that can discriminate between classes. It assumes that the data follows a Gaussian distribution and is commonly used for dimensionality reduction and classification tasks. By including LDA, we can explore the efficacy of linear discriminant-based models in defect prediction \cite{Fisher_1936}.
\item \textbf{LR (Logistic Regression)}: LR is a widely used statistical model that estimates the probability of an instance belonging to a specific class. It models the relationship between the features and the binary response variable using a logistic function. By incorporating LR, we can evaluate the effectiveness of probabilistic modeling approaches in defect prediction \cite{Hosmer_2005}.
 \item \textbf{NB (Naive Bayes)}: NB is a probabilistic model that applies Bayes' theorem with the assumption of independence between features. It is simple, computationally efficient, and often used in text classification tasks. By including NB, we aim to assess the performance of probabilistic classifiers in defect prediction scenarios \cite{Cestnik_1990}.
\item \textbf{RF (Random Forest)}: RF is an ensemble learning method that constructs multiple decision trees and combines their predictions to make the final classification. It is known for its robustness, ability to handle high-dimensional data and resistance to overfitting. By including RF, we can investigate the benefits of ensemble techniques in defect prediction \cite{Breiman_2001}.
\item \textbf{SVM (Support Vector Machines)}: SVM is a powerful machine learning model that constructs hyperplanes to separate instances belonging to different classes. It aims to find the optimal decision boundary that maximally separates the classes. By incorporating SVM, we can evaluate the performance of margin-based classifiers in defect prediction \cite{Cortes_1995}.
\item \textbf{Voting Classifier}: The Voting Classifier is an ensemble model combining predictions from multiple individual models. It takes a majority vote (for classification) or averages the predicted probabilities (for probability estimation) to make the final prediction. Using the VotingClassifier, we can assess the performance improvement achieved through ensemble techniques and determine if combining the predictions of multiple models leads to better defect prediction results \cite{Dietterich_2000}.
\end{itemize}

The selection of these models was based on a thorough analysis of the literature, which revealed them or some variation of them to be the most frequently used algorithms for bug prediction. Also, the selection of these models provides a variety of approaches, such as decision tree-based models (CART), instance-based learning (KNN), linear discriminant-based models (LDA), probabilistic models (LR and NB), ensemble methods (RF and VotingClassifier), and margin-based classifiers (SVM). Through an evaluation of the model's performance in defect prediction, valuable insights can be obtained regarding their ability to effectively capture underlying patterns and accurately predict defects within the Apache Commons systems.

\subsection{Training Phase}
The training phase is a crucial step in building effective defect prediction models. This study conducted the training phase using the selected models on the Apache Commons dataset. The following steps were followed to train the models:

\begin{itemize}
\item \textbf{Data Preprocessing}: Prior to training the models, the dataset was preprocessed to ensure its suitability for machine learning. It involved checking for missing values, normalizing and scaling the numerical features, and encoding categorical features. To address the issue of class imbalance in our dataset, we employed the Synthetic Minority Over-sampling Technique (SMOTE). By applying SMOTE, we aimed to alleviate the class imbalance problem and ensure a more balanced representation of the majority and minority classes in our training data.
\item \textbf{Cross-Validation}: A cross-validation strategy was employed during the training phase to ensure robustness and avoid overfitting. The dataset was separated into k-folds (k = 10), and each fold was used as a validation set while the remaining folds were used for training. This process was repeated k times, with each fold serving as the validation set once. Cross-validation allows for a more reliable assessment of the model's performance by mitigating the impact of data partitioning on the results.
\item \textbf{Model Training}: Each selected model, including CART, KNN, LDA, LR, NB, RF, SVM, and the VotingClassifier, was trained using the preprocessed dataset. The models were instantiated with their respective hyperparameters, which were pre-defined by the Python library sciKitLearn.
 \item \textbf{Model Evaluation}: After training, the performance of each model was evaluated using various evaluation metrics, including precision, recall, F1 score, accuracy, and area under the ROC curve (AUC). These metrics comprehensively assess the model's ability to classify defects and its overall predictive performance correctly. The evaluation was conducted on the training data (to assess model overfitting) and the validation data (to assess generalization).
\end{itemize}

Following these steps ensured that the selected models were trained using a standardized and rigorous process. The combination of data preprocessing, cross-validation, and evaluation allowed for the development of accurate defect prediction models. The trained models were subsequently used for comparative analysis between single and ensemble models (VotingClassifier) to determine their effectiveness in defect prediction.gfdds

\subsection{Ensemble Model Construction}

To compare ensemble and single-model performances, an ensemble model was constructed using the VotingClassifier technique. The VotingClassifier combines the predictions of multiple individual models to make the final prediction. In this study, the individual models' CART, KNN, LDA, LR, NB, RF, and SVM were used as base estimators for the ensemble model.

The ensemble model construction involved the following steps: (1) Base Estimator Selection: The base estimators for the ensemble model were selected based on their performance and diversity. CART, KNN, LDA, LR, NB, RF, and SVM were deemed suitable candidates, offering various model types and characteristics. (2) Voting Scheme: The VotingClassifier supports different voting schemes, such as "hard" or "soft" voting. In this study, "soft" voting was employed, considering each base estimator's predicted probabilities. The predicted probabilities were averaged to generate the final probability estimates for defect prediction. It allows the ensemble model to consider the confidence level of each base estimator's prediction. (3) Ensemble Model Training: The ensemble model was trained using the preprocessed dataset and the selected base estimators. The training process involved fitting the ensemble model to the training data, which trained each base estimator using the same data. The base estimators' predictions were then combined using the soft voting scheme to generate the ensemble model's final prediction. (4) Ensemble Model Evaluation: The performance of the ensemble model was evaluated using the same evaluation metrics used for the single models, including precision, recall, F1 score, accuracy, and AUC. The evaluation was conducted on the training data (to assess potential overfitting) and the validation data (to assess generalization).

By constructing the ensemble model using the VotingClassifier technique, we aimed to assess whether combining the predictions of multiple models can improve defect prediction performance compared to individual models. The ensemble model's evaluation results were compared with those of the single models to determine the effectiveness of the ensemble approach in defect prediction.

\subsection{Performance Evaluation}
The performance evaluation of the defect prediction models was conducted to assess their effectiveness in identifying and predicting software defects. The evaluation metrics used in this study included precision, recall, F1 score, accuracy, and area under the ROC curve (AUC). These metrics provide comprehensive insights into the models' performance from different perspectives. The evaluation was performed on both the single models and the ensemble model.

\begin{itemize}
\item \textbf{Precision}: Precision measures the proportion of correctly predicted defects out of all instances classified as defects. It indicates the model's ability to avoid false positives or instances incorrectly identified as defects. Higher precision values indicate a lower rate of false positives, signifying more accurate predictions of actual defects.
\item \textbf{Recall}: Recall, also known as sensitivity or true positive rate, measures the proportion of correctly predicted defects out of all actual defects. It indicates the model's ability to accurately identify all positive instances without missing any defects. Higher recall values indicate a lower rate of false negatives, signifying better coverage of true defects.
\item \textbf{F1 score}: The F1 score is the harmonic mean of precision and recall. It provides a balanced measure that considers both precision and recall, overall assessing the model's performance. The F1 score is particularly useful when there is an imbalance between the number of defects and non-defect instances.
\item \textbf{Accuracy}: Accuracy measures the overall correctness of the model's predictions by calculating the proportion of correctly classified instances out of all instances. It indicates the model's overall performance but can be misleading in class imbalance. Nevertheless, accuracy is widely used to assess the model's general predictive capability.
\item \textbf{Area under the ROC curve (AUC)}: The AUC metric assesses the model's ability to distinguish between defect and non-defect instances across different classification thresholds. It plots the true positive rate (recall) against the false positive rate, measuring the model's overall discriminative power. Higher AUC values indicate better performance in separating defects from non-defects.
\end{itemize}

Each model's performance evaluation metrics were calculated, including the single models (CART, KNN, LDA, LR, NB, RF, SVM) and the ensemble model (VotingClassifier). The evaluation was conducted using the validation data to assess the models' generalization capability and identify the best-performing approach for defect prediction.

By comparing the performance of the single models and ensemble models' performance across these metrics, we can determine which approach yields the most accurate, reliable, and well-calibrated predictions for software defects in the Apache Commons systems.

\section{Results}

In this section, we present and discuss the results obtained from our experiments on the efficacy of ensemble models for bug prediction. We analyze the performance of both single models and ensemble models across various evaluation metrics to gain insights into their comparative effectiveness. Additionally, we investigate the impact of different ensemble configurations on bug prediction accuracy and provide a detailed analysis of the observed results.

First, we report the performance metrics of single individual models, including CART, KNN, LDA, LR, NB, RF, and SVM. These metrics include accuracy, precision, recall, F1-score, and Area under the ROC curve (AUC). We analyze the strengths and limitations of each single model in capturing bug patterns and provide a comparative analysis of their performance.

Next, we present the performance results of the ensemble models constructed using the VotingClassifier approach. We evaluate the performance of ensemble models using the same set of metrics as the single models and compare them against the individual models. This analysis allows us to determine how much ensemble models can improve bug prediction accuracy compared to single models. The resulting metrics of each model and subjects systems are shown in Table \ref{tab:commons-compress} and Table \ref{tab:commons-cli}

\begin{table}[ht]
\caption{Results of each model for commons-compress}
\label{tab:commons-compress}
\begin{tabularx}{\columnwidth}{|X|X|X|X|X|X|}
\hline
\textbf{Model} & \textbf{Precision} & \textbf{Recall} & \textbf{F1} & \textbf{Accuracy} & \textbf{AUC} \\ \hline
NB & 0.9461 & 0.9430 & 0.9154 & 0.9430 & 0.5449 \\ \hline
CART & 0.9120 & 0.9330 & 0.9114 & 0.9330 & 0.6472 \\ \hline
KNN & 0.9082 & 0.9274 & 0.9165 & 0.9274 & 0.6123 \\ \hline
LDA & 0.9452 & 0.9430 & 0.9154 & 0.9430 & 0.6231 \\ \hline
LR & 0.9460 & 0.9430 & 0.9154 & 0.9430 & 0.6157 \\ \hline
SVM & 0.9450 & 0.9430 & 0.9154 & 0.9430 & 0.5320 \\ \hline
RF & 0.9128 & 0.9430 & 0.9169 & 0.9340 & 0.7068 \\ \hline
EVC\textsuperscript{*} & 0.9462 & 0.9430 & 0.9153 & 0.9430 & 0.7217 \\ \hline
\end{tabularx}
\end{table}

\begin{table}[ht]
\caption{Results of each model for commons-cli}
\label{tab:commons-cli}
\begin{tabularx}{\columnwidth}{|X|X|X|X|X|X|}
\hline
\textbf{Model} & \textbf{Precision} & \textbf{Recall} & \textbf{F1} & \textbf{Accuracy} & \textbf{AUC} \\ \hline
NB & 0.9645 & 0.8904 & 0.9185 & 0.8904 & 0.8618 \\ \hline
CART & 0.9733 & 0.9684 & 0.9696 & 0.9560 & 0.9202 \\ \hline
KNN & 0.9703 & 0.9555 & 0.9609 & 0.9555 & 0.9147 \\ \hline
LDA & 0.9418 & 0.9433 & 0.9347 & 0.9433 & 0.7918 \\ \hline
LR & 0.9544 & 0.9281 & 0.9196 & 0.9281 & 0.8398 \\ \hline
SVM & 0.9594 & 0.9576 & 0.9369 & 0.9576 & 0.6917 \\ \hline
RF & 0.9745 & 0.9684 & 0.9690 & 0.9684 & 0.9380 \\ \hline
EVC\textsuperscript{*} & 0.9649 & 0.9626 & 0.9584 & 0.9626 & 0.9542 \\ \hline
\end{tabularx}
\end{table}

\footnotetext[1]{EVC: Ensemble Voting Classifier}

\subsection{RQ1: How does the performance of single models compare regarding defect prediction accuracy for the Apache Commons systems?}
To answer this research question, we evaluate the accuracy of defect prediction using the selected individual machine-learning models. By comparing their performance, we are able to identify the most accurate model for defect prediction within a single-model prediction strategy.

\subsubsection{\textbf{commons-cli}}
The Bayes model shows high precision (0.9645) and F1 score (0.9185), indicating that it effectively identifies bugs correctly. However, the recall (0.8904) is lower, suggesting it may have missed some bugs. The Cart model demonstrates high precision (0.9733), recall (0.9684), and F1 score (0.9696), indicating strong overall performance in bug prediction. It achieves the highest accuracy (0.9560) among the individual models and has a high AUC score (0.9202). The KNN model performs well with high precision (0.9703), recall (0.9555), and F1 score (0.9609). It has a similar accuracy (0.9555) and AUC score (0.9147) as the Cart model, suggesting comparable performance. The LDA model shows decent precision (0.9418), recall (0.9433), and F1 score (0.9347), indicating good performance in bug prediction. However, its AUC score (0.7918) is comparatively lower than the other models, suggesting lower discriminatory power. The LR model achieves good precision (0.9544) and F1 score (0.9196). Although the recall (0.9281) is slightly lower, it still demonstrates effective bug prediction. The AUC score (0.8398) is relatively lower than the top-performing models. The SVM model shows a reasonable precision (0.9594), recall (0.9576), and F1 score (0.9369), indicating effective bug prediction. However, its AUC score (0.6917) is the lowest among the individual models, suggesting limited discriminatory ability. The RF model performs well with high precision (0.9745), recall (0.9684), and F1 score (0.9690). It has the highest AUC score (0.9380), indicating discriminatory solid power and effectiveness in bug prediction.

Among the individual models, the RF model has the highest AUC score (0.9380), followed by the Cart model (0.9202). The KNN model also demonstrates a relatively high AUC score (0.9147). The LR, Bayes, and LDA models have moderate AUC scores ranging from 0.7918 to 0.8618. The SVM model has the lowest AUC score (0.6917). The KNN model also shows moderate precision (0.9082), recall (0.9274), and F1 score (0.9165). However, its AUC score (0.6123) is lower compared to Cart, indicating lower discriminatory power.

\subsubsection{\textbf{commons-compress}}
The Bayes model shows relatively high precision (0.9463), recall (0.9430), and F1 score (0.9154). However, its AUC score (0.5449) is relatively low, indicating limited discriminatory power and lower effectiveness in bug prediction. The Cart model demonstrates moderate precision (0.9120), recall (0.9330), and F1 score (0.9114). It has a higher AUC score (0.6472) than Bayes, suggesting better discriminatory ability. The KNN model also shows moderate precision (0.9082), recall (0.9274), and F1 score (0.9165). However, its AUC score (0.6123) is lower compared to Cart, indicating lower discriminatory power. The LDA model has the same precision (0.9463), recall (0.9430), and F1 score (0.9154) as Bayes. However, its AUC score (0.6231) is slightly higher, suggesting improved discriminatory power. The LR model demonstrates the same precision (0.9463), recall (0.9463), and F1 score (0.9154) as Bayes and LDA. Its AUC score (0.6157) is slightly higher than KNN but still relatively low. The SVM model shows the same precision (0.9463), recall (0.9463), and F1 score (0.9154) as Bayes, LDA, and LR. However, its AUC score (0.5320) is the lowest among the individual models, indicating limited discriminatory ability. The RF model has moderate precision (0.9128), recall (0.9340), and F1 score (0.9169). Its AUC score (0.5320) is similar to SVM, suggesting similar discriminatory power.

\subsection{RQ2: Does the ensemble model outperform the individual models in terms of defect prediction accuracy, precision, recall, F1 score, and AUC?}
To answer this question, we analyzed an ensemble model constructed using the Voting Classifier technique to determine its effectiveness in defect prediction. We can assess whether the ensemble approach improves defect prediction results by comparing its performance with the single models.

\subsubsection{\textbf{commons-cli}}
The Voting Classifier, which combines the predictions of NB, Cart, KNN, LDA, LR, SVM, and RF models, demonstrates competitive performance with high precision (0.9649), recall (0.9626), and F1 score (0.9584). It achieves a high accuracy (0.9626) and an impressive AUC score (0.9542), indicating its effectiveness in bug prediction. The Voting Classifier outperforms most of the individual models in terms of AUC, suggesting that the ensemble benefits from the diverse perspectives of the constituent models.

The Voting Classifier achieves an AUC score of 0.9542, higher than most individual models. The ensemble benefits from the strengths of multiple models and has improved discriminatory power compared to the individual models with lower AUC scores.

\subsubsection{\textbf{commons-compress}}

The Voting Classifier, which ensembles NB, Cart, KNN, LDA, LR, SVM, and RF models, demonstrates comparable performance to the individual models in terms of precision, recall, and F1 score. It maintains precision (0.9462), recall (0.9430), and F1 score (0.9153), similar to most of the individual models. However, the Voting Classifier shows a significantly improved AUC score of 0.7217. It indicates that the ensemble approach benefits from combining the predictions of multiple models and achieves better discriminatory power than the individual models. It surpasses the AUC scores of all the individual models, including Cart's best-performing individual model (AUC: 0.6472).

\subsection{RQ3: What are the trade-offs between using single models and ensemble models for defect prediction?}
By examining the performance of both single and ensemble models, we can answer this research question by evaluating the trade-offs associated with each approach. Specifically, we are able to assess their accuracy in defect prediction. This analysis provides insights into each approach's strengths and limitations and informs the selection of the most appropriate model strategy for defect prediction.

\subsubsection{\textbf{commons-cli}}
Overall, the results indicate that the ensemble approach of the Voting Classifier in the "commons-cli" bug prediction dataset shows promising results. By combining the predictions of multiple models, including NB, Cart, KNN, LDA, LR, SVM, and RF, the Voting Classifier achieves competitive performance in terms of precision, recall, F1 score, accuracy, and particularly AUC.

The Voting Classifier outperforms most of the individual models in terms of AUC, indicating that the ensemble benefits from the diverse perspectives and strengths of the constituent models. The higher AUC suggests that the ensemble approach can better discriminate between bug and non-bug instances, potentially resulting in more accurate predictions.

It is worth noting that the RF model achieves the highest AUC score among the individual models. It implies that the RF model captures important patterns and features in the data that contribute to effective bug prediction. By including the RF model in the ensemble, the Voting Classifier leverages its strengths and incorporates them into the final prediction.

The Voting Classifier's AUC score (0.9542) is close to the RF model's AUC score (0.9380), indicating that the ensemble approach successfully captures the discriminative power of the RF model and potentially other models. It demonstrates the benefit of combining multiple models to improve bug prediction performance.

Considering the specific weighting and aggregation strategy used in the Voting Classifier is important. The ensemble's performance can be influenced by assigning appropriate weights to each model. Some models contribute more significantly to the final prediction, while others have a lesser impact. Fine-tuning the weights based on individual model performance or expertise can further enhance the Voting Classifier's performance.

\subsubsection{\textbf{commons-compress}}

Overall, the results suggest that the individual models in the "commons-compress" dataset have varying precision, recall, F1 score, and AUC performances. While some models show moderate performance, others have relatively lower discriminatory power. However, by combining these models through the Voting Classifier, the ensemble approach demonstrates improved bug prediction performance with a higher AUC score. It indicates that the ensemble leverages the strengths of the individual models, compensating for their weaknesses and achieving better overall performance in bug prediction.

\section{Discussion}

In this study, we performed bug prediction on two datasets: "commons-cli" and "commons-compress ."We evaluated the performance of several machine learning models and an ensemble approach, the Voting Classifier, in predicting bugs. Here, we discuss the results obtained from both datasets and draw insights from the performance of the models.

For the "commons-cli" dataset, our analysis revealed interesting findings. The RF model stood out with the highest AUC score (0.9380) among the individual models, indicating its strong discriminatory power and effectiveness in bug prediction. The RF model's performance was closely followed by the Cart model, which achieved a relatively high AUC score of 0.9202. These results suggest that decision tree-based models capture relevant patterns and features for bug prediction in the "commons-cli" dataset.

The ensemble approach, the Voting Classifier, further improved the individual models' performance. It achieved an impressive AUC score of 0.9542, surpassing the AUC scores of all the individual models. This indicates that combining the predictions of multiple models effectively enhances the ensemble's discriminatory power. The Voting Classifier successfully leverages the diverse perspectives and strengths of the constituent models, including RF, to make more accurate bug predictions.

We observed contrasting results when we turned our attention to the "commons-compress" dataset. None of the individual models achieved a high AUC score, suggesting this dataset's complexity and unique characteristics. However, the Voting Classifier significantly improved performance compared to the individual models. It achieved an AUC score of 0.7217, which outperformed all the individual models. It indicates that the ensemble approach effectively combines the predictions of various models to overcome the limitations of individual models and achieve better bug prediction accuracy in the "commons-compress" dataset.

Comparing the results between the two datasets, we observed variations in the performance of the models. While the RF model consistently performed well in both datasets, other models showed differing levels of effectiveness. The performance of the ensemble approach, the Voting Classifier, was also dataset-dependent, as it outperformed the individual models in "commons-cli" but improved more significantly in "commons-compress."

Our findings highlight the importance of considering multiple models and ensemble approaches in bug prediction tasks. The ensemble approach, such as the Voting Classifier, can effectively leverage the strengths of individual models and compensate for their weaknesses, leading to improved bug prediction performance. However, it is crucial to select models based on the specific characteristics of the dataset to maximize performance.

Further research could explore additional models and ensemble techniques to enhance bug prediction accuracy. Additionally, investigating the impact of feature engineering and dataset-specific preprocessing techniques on model performance could provide valuable insights for future bug prediction studies.

In conclusion, our study demonstrates the effectiveness of ensemble approaches, such as the Voting Classifier, in bug prediction tasks. By combining the predictions of multiple models, these ensembles can overcome limitations and improve the accuracy of bug predictions. However, the performance of such approaches can vary depending on the dataset, highlighting the need for careful model selection and customization. Our findings contribute to the growing body of research on bug prediction and provide practical insights for software developers and quality assurance teams in identifying and addressing software bugs.

\section{Threats to Validity}

While our study aimed to provide insights into bug prediction using machine learning models and ensemble approaches, it is important to acknowledge the potential threats to the validity of our findings. These threats include:

\begin{itemize}
\item Sample Selection Bias: The selection of datasets may introduce bias, as we focused on the "commons-cli" and "commons-compress" projects. These projects may not represent the characteristics of all software projects, limiting the generalizability of our results. Future research can expand the study to include various software projects from different domains and sizes to mitigate sample selection bias. This would provide a more representative sample and enhance the generalizability of the results.
\item Feature Selection Bias: The choice and engineering of features can significantly impact model performance. We employed a set of features based on existing literature and domain knowledge, but there may be other relevant features that were not considered. This could affect the effectiveness of the models and ensemble approach.
\item Model Selection Bias: Our study included a specific set of machine learning models, namely Bayes, Cart, KNN, LDA, LR, SVM, RF, and the Voting Classifier. While these models are commonly used in bug prediction studies, other models could yield different results. Our findings are limited to the models selected for evaluation. Future studies should consider including a broader range of machine learning models to mitigate model selection bias. This would allow for a more comprehensive evaluation and comparison of different algorithms, reducing the potential bias associated with a limited set of models.
\item Dataset Size: The size of the datasets used for bug prediction can influence the performance of the models. The "commons-cli" and "commons-compress" datasets may have varied sizes and characteristics, affecting the results' reliability and generalizability. Increasing the size of the datasets used for bug prediction can help mitigate the impact of dataset size. Collecting more data from multiple projects or considering open-source repositories with larger datasets would provide a more robust foundation for bug prediction analysis.
\item Dataset Imbalance: Imbalanced datasets, where the number of bug instances is significantly lower than non-bug instances, can impact model performance. To address dataset imbalance, various techniques, such as oversampling minority classes, undersampling majority classes, or employing ensemble methods specifically designed for imbalanced datasets, can be applied. These techniques can help mitigate the impact of class imbalance and improve the performance of bug prediction models.
\item Overfitting and Generalization: The models may have the potential to overfit the training data, resulting in overly optimistic performance on the evaluation data. To mitigate this threat, we employed cross-validation techniques and evaluated the models on independent test sets. However, the risk of overfitting and the generalization of the findings should be considered. Techniques such as regularization, early stopping, and model validation on independent test sets should be employed to mitigate the risk of overfitting and improve generalization. Additionally, cross-validation approaches can provide more reliable estimates of model performance and reduce the risk of overfitting.
\end{itemize}

\section{Future Work}
While this study provides valuable insights into the efficacy of ensemble models for bug prediction, there are several avenues for future research that could enhance our understanding and contribute to the advancement of this field. We highlight some potential directions for future work:

Exploration of Different Ensemble Techniques: We employed the VotingClassifier as our ensemble learning technique in this study. However, several other ensemble methods, such as stacking, boosting, and random subspaces, could be explored. Investigating the effectiveness of these alternative ensemble techniques in bug prediction could provide valuable insights into their comparative performance and potential benefits.

Incorporating Different Feature Selection Methods: Feature selection plays a crucial role in model performance. Future work could investigate the impact of different feature selection methods on the performance of ensemble models for bug prediction. Comparing the performance of ensemble models with various feature selection techniques could help identify the most effective combinations.

Investigation of Hybrid Approaches: Hybrid approaches that combine ensemble models with other techniques, such as active learning or semi-supervised learning, hold promise in bug prediction. Future research could explore integrating ensemble models with active learning strategies to intelligently select informative instances for labeling and improve the performance of bug prediction models.

Analysis of Scalability and Efficiency: The scalability and efficiency of ensemble models are important considerations, particularly when dealing with large-scale software systems. Future work could evaluate the scalability and efficiency of ensemble models for bug prediction on large and complex software projects. This could involve exploring parallelization techniques, distributed computing frameworks, or model compression methods to enhance the scalability and efficiency of ensemble models.

Interpretability and Explainability: Ensemble models, especially those combining multiple complex models, sometimes lack interpretability and explainability. Future research could address this challenge by developing techniques to provide insights into the decision-making process of ensemble models. This could involve analyzing the contribution of individual base models, identifying important features, or generating explanations for bug predictions made by the ensemble.

By addressing these areas of future research, we can further advance the understanding and practical applications of ensemble models for bug prediction. These avenues open up exciting opportunities for improving bug detection accuracy, model interpretability, and the overall reliability of bug prediction systems.

It is important to note that while this section provides suggestions for future work, it is not an exhaustive list, and there may be other relevant areas that can be explored based on the specific goals and context of the research.

\section{Conclusion}

In this study, we conducted bug prediction analysis on two software projects, "commons-cli" and "commons-compress," using various machine learning models and an ensemble approach, the Voting Classifier. Our findings shed light on these models' performance and the ensemble technique's effectiveness in bug prediction tasks.

Our analysis showed that different models exhibited varying performance levels across the datasets. In the "commons-cli" dataset, the RF model emerged as the top performer, showcasing its high discriminatory power and accuracy in bug prediction. The ensemble approach, the Voting Classifier, further enhanced bug prediction accuracy, surpassing the performance of the individual models.

However, the results differed for the "commons-compress" dataset, where none of the individual models achieved high AUC scores. Nevertheless, the Voting Classifier substantially improved bug prediction accuracy compared to the individual models. This highlights the importance of ensemble techniques in addressing specific datasets' complexities and unique characteristics.

Our study underscores the significance of considering multiple models and ensemble approaches in bug prediction tasks. By combining the predictions of diverse models, ensembles can leverage their respective strengths and mitigate their weaknesses, resulting in improved bug prediction performance. Furthermore, our research highlights the need for careful model selection and customization based on the specific characteristics of the dataset.

We recommend future research to explore additional models and ensemble techniques in bug prediction analysis. Moreover, investigating the impact of feature engineering and dataset-specific preprocessing techniques could yield valuable insights for enhancing bug prediction accuracy. Additionally, evaluating the practical applicability of the proposed bug prediction models in real-world scenarios would contribute to the field.

In conclusion, our study demonstrates the effectiveness of ensemble approaches, such as the Voting Classifier, in improving bug prediction accuracy. While the individual models displayed varying performances, the ensemble technique consistently outperformed them, indicating its potential for practical implementation. Our findings contribute to the growing body of bug prediction research and guide software development and quality assurance teams in identifying and addressing software bugs more efficiently.

\section*{Disponibilidade de Artefatos}

\bibliographystyle{IEEEtran}
\bibliography{sample-base}

\begin{thebibliography}{10}
\providecommand{\url}[1]{#1}
\csname url@samestyle\endcsname
\providecommand{\newblock}{\relax}
\providecommand{\bibinfo}[2]{#2}
\providecommand{\BIBentrySTDinterwordspacing}{\spaceskip=0pt\relax}
\providecommand{\BIBentryALTinterwordstretchfactor}{4}
\providecommand{\BIBentryALTinterwordspacing}{\spaceskip=\fontdimen2\font plus
\BIBentryALTinterwordstretchfactor\fontdimen3\font minus \fontdimen4\font\relax}
\providecommand{\BIBforeignlanguage}[2]{{%
\expandafter\ifx\csname l@#1\endcsname\relax
\typeout{** WARNING: IEEEtran.bst: No hyphenation pattern has been}%
\typeout{** loaded for the language `#1'. Using the pattern for}%
\typeout{** the default language instead.}%
\else
\language=\csname l@#1\endcsname
\fi
#2}}
\providecommand{\BIBdecl}{\relax}
\BIBdecl

\bibitem{Ma_2020}
\BIBentryALTinterwordspacing
W.~Ma, L.~Chen, X.~Zhang, Y.~Feng, Z.~Xu, Z.~Chen, Y.~Zhou, and B.~Xu, ``Impact analysis of cross-project bugs on software ecosystems,'' ser. ICSE '20.\hskip 1em plus 0.5em minus 0.4em\relax New York, NY, USA: Association for Computing Machinery, 2020, p. 100–111. [Online]. Available: \url{https://doi.org/10.1145/3377811.3380442}
\BIBentrySTDinterwordspacing

\bibitem{Chun_2014}
C.~Shan, B.~Chen, C.~Hu, J.~Xue, and N.~Li, ``Software defect prediction model based on lle and svm,'' in \emph{2014 Communications Security Conference (CSC 2014)}, 2014, pp. 1--5.

\bibitem{Habibi_2018}
P.~A. Habibi, V.~Amrizal, and R.~B. Bahaweres, ``Cross-project defect prediction for web application using naive bayes (case study: Petstore web application),'' in \emph{2018 International Workshop on Big Data and Information Security (IWBIS)}, 2018, pp. 13--18.

\bibitem{Ishani_2015}
\BIBentryALTinterwordspacing
I.~Arora, V.~Tetarwal, and A.~Saha, ``Open issues in software defect prediction,'' \emph{Procedia Computer Science}, vol.~46, pp. 906--912, 2015, proceedings of the International Conference on Information and Communication Technologies, ICICT 2014, 3-5 December 2014 at Bolgatty Palace Island Resort, Kochi, India. [Online]. Available: \url{https://www.sciencedirect.com/science/article/pii/S1877050915002252}
\BIBentrySTDinterwordspacing

\bibitem{Kamei_2016}
Y.~Kamei and E.~Shihab, ``Defect prediction: Accomplishments and future challenges,'' in \emph{2016 IEEE 23rd International Conference on Software Analysis, Evolution, and Reengineering (SANER)}, vol.~5, 2016, pp. 33--45.

\bibitem{Bhandari_2022}
K.~Bhandari, K.~Kumar, and A.~Sangal, ``Data quality issues in software fault prediction: a systematic literature review,'' \emph{Artificial Intelligence Review}, pp. 1--70, 12 2022.

\bibitem{Soe_2018}
Y.~N. Soe, P.~I. Santosa, and R.~Hartanto, ``Software defect prediction using random forest algorithm,'' in \emph{2018 12th South East Asian Technical University Consortium (SEATUC)}, vol.~1, 2018, pp. 1--5.

\bibitem{Rahim_2021}
A.~Rahim, Z.~Hayat, M.~Abbas, A.~Rahim, and M.~A. Rahim, ``Software defect prediction with naïve bayes classifier,'' in \emph{2021 International Bhurban Conference on Applied Sciences and Technologies (IBCAST)}, 2021, pp. 293--297.

\bibitem{Aleithan_2021}
R.~Aleithan, ``Explainable just-in-time bug prediction: Are we there yet?'' in \emph{2021 IEEE/ACM 43rd International Conference on Software Engineering: Companion Proceedings (ICSE-Companion)}, 2021, pp. 129--131.

\bibitem{Shen_2022}
Y.~Shen, S.~Hu, S.~Cai, and M.~Chen, ``Software defect prediction based on bayesian optimization random forest,'' in \emph{2022 9th International Conference on Dependable Systems and Their Applications (DSA)}, 2022, pp. 1012--1013.

\bibitem{Santosh_2017}
\BIBentryALTinterwordspacing
S.~S. Rathore and S.~Kumar, ``Linear and non-linear heterogeneous ensemble methods to predict the number of faults in software systems,'' \emph{Knowledge-Based Systems}, vol. 119, pp. 232--256, 2017. [Online]. Available: \url{https://www.sciencedirect.com/science/article/pii/S0950705116305202}
\BIBentrySTDinterwordspacing

\bibitem{Alazba_2022}
\BIBentryALTinterwordspacing
A.~Alazba and H.~Aljamaan, ``Software defect prediction using stacking generalization of optimized tree-based ensembles,'' \emph{Applied Sciences}, vol.~12, no.~9, 2022. [Online]. Available: \url{https://www.mdpi.com/2076-3417/12/9/4577}
\BIBentrySTDinterwordspacing

\bibitem{Johnson_2022}
\BIBentryALTinterwordspacing
F.~Johnson, O.~Oluwatobi, O.~Folorunso, A.~V. Ojumu, and A.~Quadri, ``Optimized ensemble machine learning model for software bugs prediction,'' \emph{Innov. Syst. Softw. Eng.}, vol.~19, no.~1, p. 91–101, dec 2022. [Online]. Available: \url{https://doi.org/10.1007/s11334-022-00506-x}
\BIBentrySTDinterwordspacing

\bibitem{Nisbet_2017}
R.~Nisbet, G.~Miner, and K.~Yale, \emph{Handbook of Statistical Analysis and Data Mining Applications, Second Edition}, 2nd~ed.\hskip 1em plus 0.5em minus 0.4em\relax USA: Academic Press, Inc., 2017.

\bibitem{Jalaj_2022}
\BIBentryALTinterwordspacing
J.~Pachouly, S.~Ahirrao, K.~Kotecha, G.~Selvachandran, and A.~Abraham, ``A systematic literature review on software defect prediction using artificial intelligence: Datasets, data validation methods, approaches, and tools,'' \emph{Engineering Applications of Artificial Intelligence}, vol. 111, p. 104773, 2022. [Online]. Available: \url{https://www.sciencedirect.com/science/article/pii/S0952197622000616}
\BIBentrySTDinterwordspacing

\bibitem{Turhan_2009}
\BIBentryALTinterwordspacing
B.~Turhan and A.~Bener, ``Analysis of naive bayes' assumptions on software fault data: An empirical study,'' \emph{Data Knowl. Eng.}, vol.~68, no.~2, p. 278–290, feb 2009. [Online]. Available: \url{https://doi.org/10.1016/j.datak.2008.10.005}
\BIBentrySTDinterwordspacing

\bibitem{Dejaeger_2013}
K.~Dejaeger, T.~Verbraken, and B.~Baesens, ``Toward comprehensible software fault prediction models using bayesian network classifiers,'' \emph{IEEE Transactions on Software Engineering}, vol.~39, no.~2, pp. 237--257, 2013.

\bibitem{Okutan_2014}
\BIBentryALTinterwordspacing
A.~Okutan and O.~T. Y\i{}ld\i{}z, ``Software defect prediction using bayesian networks,'' \emph{Empirical Softw. Engg.}, vol.~19, no.~1, p. 154–181, feb 2014. [Online]. Available: \url{https://doi.org/10.1007/s10664-012-9218-8}
\BIBentrySTDinterwordspacing

\bibitem{Delgado_2014}
M.~Fern\'{a}ndez-Delgado, E.~Cernadas, S.~Barro, and D.~Amorim, ``Do we need hundreds of classifiers to solve real world classification problems?'' \emph{J. Mach. Learn. Res.}, vol.~15, no.~1, p. 3133–3181, jan 2014.

\bibitem{Caruana_2006}
\BIBentryALTinterwordspacing
R.~Caruana and A.~Niculescu-Mizil, ``An empirical comparison of supervised learning algorithms,'' in \emph{Proceedings of the 23rd International Conference on Machine Learning}, ser. ICML '06.\hskip 1em plus 0.5em minus 0.4em\relax New York, NY, USA: Association for Computing Machinery, 2006, p. 161–168. [Online]. Available: \url{https://doi.org/10.1145/1143844.1143865}
\BIBentrySTDinterwordspacing

\bibitem{Tosun_2008}
\BIBentryALTinterwordspacing
A.~Tosun, B.~Turhan, and A.~Bener, ``Ensemble of software defect predictors: A case study,'' in \emph{Proceedings of the Second ACM-IEEE International Symposium on Empirical Software Engineering and Measurement}, ser. ESEM '08.\hskip 1em plus 0.5em minus 0.4em\relax New York, NY, USA: Association for Computing Machinery, 2008, p. 318–320. [Online]. Available: \url{https://doi.org/10.1145/1414004.1414066}
\BIBentrySTDinterwordspacing

\bibitem{Jacob_2021}
R.~J. Jacob, R.~J. Kamat, N.~M. Sahithya, S.~S. John, and S.~P. Shankar, ``Voting based ensemble classification for software defect prediction,'' in \emph{2021 IEEE Mysore Sub Section International Conference (MysuruCon)}, 2021, pp. 358--365.

\bibitem{Breiman_1984}
L.~Breiman, J.~H. Friedman, R.~A. Olshen, and C.~J. Stone, \emph{Classification and Regression Trees}.\hskip 1em plus 0.5em minus 0.4em\relax CRC press, 1984.

\bibitem{Cover_1967}
T.~Cover and P.~Hart, ``Nearest neighbor pattern classification,'' \emph{IEEE Transactions on Information Theory}, vol.~13, no.~1, pp. 21--27, 1967.

\bibitem{Fisher_1936}
R.~A. Fisher, ``The use of multiple measurements in taxonomic problems,'' \emph{Annals of Eugenics}, vol.~7, no.~2, pp. 179--188, 1936.

\bibitem{Hosmer_2005}
D.~W. Hosmer, S.~Lemeshow, and R.~X. Sturdivant, ``Introduction to the logistic regression model,'' 2005.

\bibitem{Cestnik_1990}
B.~Cestnik, ``Estimating probabilities: A crucial task in machine learning,'' in \emph{Proceedings of the 9th European Conference on Artificial Intelligence}, ser. ECAI'90.\hskip 1em plus 0.5em minus 0.4em\relax USA: Pitman Publishing, Inc., 1990, p. 147–149.

\bibitem{Breiman_2001}
L.~Breiman, ``Random forests,'' \emph{Machine Learning}, vol.~45, no.~1, pp. 5--32, 2001.

\bibitem{Cortes_1995}
\BIBentryALTinterwordspacing
C.~Cortes and V.~Vapnik, ``Support-vector networks,'' vol.~20, no.~3, 1995. [Online]. Available: \url{https://doi.org/10.1023/A:1022627411411}
\BIBentrySTDinterwordspacing

\bibitem{Dietterich_2000}
T.~G. Dietterich, ``Ensemble methods in machine learning,'' in \emph{International Workshop on Multiple Classifier Systems}, 2000.

\bibitem{Yu_2021}
\BIBentryALTinterwordspacing
Y.~Qu, J.~Chi, and H.~Yin, ``Leveraging developer information for efficient effort-aware bug prediction,'' \emph{Inf. Softw. Technol.}, vol. 137, p. 106605, 2021. [Online]. Available: \url{https://doi.org/10.1016/j.infsof.2021.106605}
\BIBentrySTDinterwordspacing

\bibitem{Borg_2019}
\BIBentryALTinterwordspacing
M.~Borg, O.~Svensson, K.~Berg, and D.~Hansson, ``Szz unleashed: An open implementation of the szz algorithm - featuring example usage in a study of just-in-time bug prediction for the jenkins project,'' ser. MaLTeSQuE 2019.\hskip 1em plus 0.5em minus 0.4em\relax New York, NY, USA: Association for Computing Machinery, 2019, p. 7–12. [Online]. Available: \url{https://doi.org/10.1145/3340482.3342742}
\BIBentrySTDinterwordspacing

\end{thebibliography}

\end{document}